\documentstyle[preprint,prl,aps,graphicx]{revtex}
\begin{document}
%\twocolumn[\hsize\textwidth\columnwidth\hsize\csname
%@twocolumnfalse\endcsname

%\setlength{\topmargin}{-0.5cm}
\setlength{\topmargin}{-1.5cm}
\setlength{\headsep}{1.6cm}
\setlength{\evensidemargin}{-1.cm}
\setlength{\oddsidemargin}{-0.cm}
\setlength{\textheight}{23cm}
\setlength{\textwidth}{18cm}
\newcommand \be{\begin{equation}}
\newcommand \ba{\begin{eqnarray}}
\newcommand \ee{\end{equation}}
\newcommand \ea{\end{eqnarray}}
\newcommand{\lp}{\left(}
\newcommand{\rp}{\right)}

%% Newly defined %%%%%%%%%%%%%%%%%%%%%%%%%%%
%\newcommand \kcmn[1]{{\sf [{#1}]}  \marginpar{\bf [$\bullet$]}}
% variable y & b
\newcommand{\yo}{y_{1}}
\newcommand{\yt}{y_{2}}
\newcommand{\vy}{{\bf y}}
\newcommand{\vyz}{{\bf y}_{0}}
\newcommand{\tz}{t_{0}}

%% Title page %%%%%%%%%%%%%%%%%%%%%%%%%%%%%%%%%%%%%%%
\begin{center}
\LARGE Theory of self-similar oscillatory finite-time singularities
in Finance, Population and Rupture
\end{center}
\bigskip
\begin{center}
\Large Didier  Sornette$^{\mbox{\ref{ess},\ref{lpec}}}$ and
Kayo Ide$^{\mbox{\ref{das}}}$\\
\end{center}
\begin{center}
Institute of Geophysics and Planetary Physics\\ 
University of California, Los Angeles\\ 
Los Angeles, CA 90095-1567\label{igpp}
\end{center}

\bigskip
%\begin{singlespace}
\begin{enumerate}

\item Also at the Department of Earth and Space Sciences, UCLA\label{ess}

\item Also at the Laboratoire de Physique de la Mati\`ere Condens\'ee, CNRS UMR
6622 and Universit\'e de Nice-Sophia Antipolis, 06108 Nice Cedex 2,
France\label{lpec}

\item Also at the Department of Atmospheric Sciences, UCLA\label{das}

\end{enumerate}
%\end{singlespace}

\begin{abstract}

We present a simple two-dimensional dynamical system reaching
a singularity in finite time decorated by accelerating oscillations due to
the interplay between  nonlinear positive feedback and 
reversal in the inertia. This provides a fundamental equation for the 
dynamics of (1) stock market prices in the presence of 
nonlinear trend-followers and nonlinear value investors, (2) 
the world human population with a competition between a
population-dependent growth rate and a nonlinear 
dependence on a finite carrying capacity and
(3) the failure of a material
subject to a time-varying stress with a competition between positive
geometrical feedback on the damage variable and nonlinear healing.
The rich fractal scaling properties of the dynamics are traced back to
the self-similar spiral structure in phase space
unfolding around an unstable spiral point at the origin.

\end{abstract}
\vskip 0.5cm

%\narrowtext
%\twocolumn[...text...]

Singularities play an important role in the physics of phase transitions
as well as in signatures of positive feedbacks in dynamical systems,
with examples in the Euler
equations of inviscid fluids \cite{Pumiersiggia}, in
vortex collapse of systems of point vortices, in the equations of
General Relativity coupled to a mass field leading to the formation of
black holes \cite{Choptuik}, in models of micro-organisms aggregating to form
fruiting bodies \cite{Rascle}, in models of material failure 
\cite{critrupt}, of
earthquakes \cite{earthquake} and of stock market crashes \cite{crash}.

The continuous scale invariance usually associated 
with a singularity can be partially broken into a weaker
symmetry, called discrete scale invariance (DSI), 
according to which the  self-similarity
holds only for integer powers of a specific factor $\lambda$ \cite{reviewsor}.
The hallmark of this DSI is the transformation of 
the power law into an oscillatory singularity, with log-periodic
oscillations decorating the overall power law acceleration towards the
singularity. 
Such log-periodic power laws have been documented for many systems 
such as with a built-in geometrical hierarchy, 
as the result of a cascade of ultra-violet instabilities in growth 
processes and rupture, in deterministic dynamical systems,
in response functions of spin systems with quenched disorder, 
in spinodal decomposition of binary mixtures in uniform shear flow, etc.
(see \cite{reviewsor,BookSor} and references therein).

Here, we introduce a general dynamical mechanism for a
finite-time singularity with self-similar oscillatory behavior,
based on the interplay between 
nonlinear positive feedback and reversal in the inertia:
\ba
{d y_1 \over dt} &=& y_2~,  \label{dyn1}\\
{d y_2 \over dt} &=& y_2 |y_2|^{m-1} - \gamma y_1 |y_1|^{n-1}~.
\label{dyn2}
\ea
This model is  motivated by and derived from the dynamics of
stock market prices, of
the world human population and of material failure  as we shall see
below (see \cite{idesor} for detailed derivations).
Our analysis of (\ref{dyn1}, \ref{dyn2})
offers a fundamental understanding of the observed
interplay between accelerating growth and accelerating (log-periodic)
oscillations previously documented in speculative bubbles preceding
large crashes \cite{crash,nasdaq},   
the world human population \cite{Kapitza,JohSorgreat},
and  time-to-failure analysis of material rupture
\cite{critrupt} (and references therein).

{\bf Stock market price dynamics}: 
the heterogeneous behavior of agents has recently been shown to be a
crucial ingredient to account for the  
complexity of financial time series (see \cite{Luxnature} and
references therein).
Typically, ``value investors'' track the fundamental price $p_f$ of a
given stock placing 
investment orders of (algebraic) size $\Omega_{\rm value}(t)$
while ``technical analysts'' use trend following strategies to place
investment orders of size $\Omega_{\rm tech}(t)$. 
The balance between supply and demand
determines the price variation from $p(t)$ to
$p(t+\delta t)$ over the time interval $\delta t$
according to $\ln p(t+\delta t) - \ln p(t)  = {1 \over L} 
[\Omega_{\rm value}(t) + \Omega_{\rm tech}(t)]$ \cite{Farmer}
where $L>0$ is a market depth. We postulate the nonlinear dependence
$\Omega_{\rm value}(t) = -c ~\ln [p(t)/p_f] ~ |\ln [p(t)/ p_f]|^{n-1}$, 
where $n>1$ and $c>0$ is a constant. The case $n=1$ retrieves an 
ingredient of previous models \cite{Farmer,Bouchaudcont}.
According to textbook economics, $p_f$ is determined by the 
discounted expected future dividends whose estimation is very sensitive to
the forecast of their growth rate and of the interest rate, leading to 
large uncertainties in $p_f$. As a consequence, a trader
trying to track fundamental value has no incentive to react when she feels
that the deviation is small since this deviation is more or less within
the noise. Only when the departure of price from fundamental value
becomes relatively large will the trader act. The 
exponent $n>1$ precisely accounts for this effect.
The second class of investors follow
strategies that are positively related to past price moves. This can
be captured by the following contribution to the order size:
$\Omega_{\rm tech}(t) 
 = a_1[\ln p(t) - \ln p(t-\delta t)] +a_2 [\ln p(t) - \ln p(t-\delta t)]
|\ln p(t) - \ln p(t-\delta t)|^{m-1}$ with $a_1>0$ and $a_2>0$.
The choice $m>1$ means that trend-following strategies tend to
under-react  for small price changes and over-react for large ones.
Posing $y_1(t) = \ln [p(t)/p_f]$, we expand
the equation of balance between supply and demand 
as a Taylor series in powers of $\delta t$ and get
\be
(\delta t)^2 {d^2 x \over dt^2} = - \left[1-{a_1 \over L}\right]~\delta t~ {dx \over dt} +
{a_2 (\delta t)^{m}  \over L} {dx \over dt}
|{dx \over dt}|^{m-1} - {c \over L} x(t)  |x(t)|^{n-1}~+~{\cal O}[(\delta t)^3]~,
\ee
where ${\cal O}[(\delta t)^3]$ represents a term of the order of $(\delta t)^3$.
This equation is a generalization of the model of
Pandey and Stauffer \cite{PandeyStauffer}, by allowing nonlinear
positive feedback $m>1$ of the trend-following strategies.
The theory becomes critical when the ``mass'' term vanishes, i.e.,
when $a_1=L$. Rescaling $t$ and $y_1$ by $\alpha$ and posing 
$\yt=d\yo/dt$ and $\gamma=\alpha^{-(n+1)}c /L(\delta t)^2$ 
where $\alpha\equiv a_2 (\delta t)^{m-2} /L$, 
we obtain (\ref{dyn1},\ref{dyn2}).

{\bf Population dynamics}:
the logistic equation corrects Malthus' exponential growth model
by assuming that the population $p(t)$ cannot growth beyond 
the earth carrying capacity $K$:
${d p \over dt} =  \sigma_0 p(t) \left[K - p(t)\right]$ 
where $\sigma_{0}$ controls the amplitude of the nonlinear saturation term
(see \cite{Cohenscience} and references therein). However, it is now
understood that $K$ is not a constant but
increases with $p(t)$ due to technological progress such as the
use of tools and fire, the development of agriculture, the use of fossil fuels,
fertilizers {\it etc.} as well an expansion into new habitats and the removal
of limiting factors by the development of vaccines, pesticides, antibiotics.
If $K(t)$ grows faster than $p(t)$, then $p(t)$
explodes to infinity after a finite time creating a singularity
due to the corresponding growth of the growth rate
$\sigma \equiv d\ln p/dt$, leading to
${d\sigma \over dt} \propto \sigma^2$ \cite{JohSorgreat}.
We now generalize it by assuming a nonlinear saturation: 
${d\sigma \over dt} = \alpha \sigma |\sigma|^{m-1} 
 - \gamma \ln (p/K_{\infty}) |\ln (p/K_{\infty})|^{n-1}$
where $\alpha$ and $\gamma>0$ measure the effect of feedback
and reversal.
Apart from the absolute value, the first term in the r.h.s. is the previous
nonlinear growth of the growth rate. The novel second term favors a restoration
of the population $p(t)$ to the asymptotic carrying capacity $K_{\infty}$. 
The effective cumulative growth rate
$\ln (p/K_{\infty})$ is the natural variable to describe the attraction to 
$K_{\infty})$.
The nonlinear restoring exponent $n>1$ captures
the many nonlinear (often quasi-threshold) feedback mechanisms acting 
on population dynamics. 
Defining variables $(y_1,y_2)=(\alpha \ln (p/K_{\infty}),\sigma)$ 
and rescaling $t$ by $\alpha$ lead to  (\ref{dyn1},\ref{dyn2}).

{\bf Rupture of materials with competing damage and healing}:
a standard model of damage rupture \cite{Krajci} consists in a rod
subjected to uniaxial tension by a constant applied axial force $P$.
The undamaged cross section $A(t)$ of the rod is assumed to be
a function of time but is independent of the axial coordinate. 
The considered viscous
deformation is assumed to be isochoric: $A_0 L_0 = A(t) L(t) =$ constant
at all times, where $L(t)$ is the length of the rod. The creep strain rate 
${d \epsilon_c \over dt} = {1 \over L}{dL \over dt} 
 = -{1 \over A}{dA  \over dt}$ is assumed to follow  Norton's law:
${d \epsilon_c \over dt} = C \sigma^{\mu}$ where $\sigma = P/A$
is the rod cross section with $C>0$ and $\mu>0$ 
Eliminating $d \epsilon_c / dt$ leads to
$A^{\mu - 1} dA/dt = - CP^{\mu}$,
%%$A(t) = P (\mu C)^{1/\mu} (t_c-t)^{1/\mu}$, where
%%the critical failure time $t_c$ is given by the equation 
%%$P (\mu C t_c)^{1/\mu} = A_0$.
%%The rod cross section thus vanishes in a finite time $t_c$ and
%%as a consequence the stress diverges as the time $t$ goes to the 
%%critical time $t_c$ as $\sigma = P/A  = (\mu C)^{-1/\mu} (t_c-t)^{-1/\mu}$
and hence ${d \sigma \over dt} = C \sigma^{\mu+1}$.
Physically, this results from a geometrical feedback of the undamaged 
area on the stress and vice-versa.

We generalize this model by adding healing as well as a
strain-dependent loading: 
${d \sigma \over dt} = \alpha \sigma |\sigma|^{m-1} 
 - \gamma  \epsilon_c |\epsilon_c|^{n-1}$.
The first term in the right-hand-side is identical to
previous geometrical positive feedback for $m=\mu-1$. We relax
this correspondence and allow $m>1$ to be arbitrary. The addition of 
the second term introduces the  physical ingredient that damage can be
reversible. 
Large deformations can enhance healing which increases the undamage
area and thus 
decrease the effective stress within the material.
The model is completed by using again Norton's law but with exponent $m'$:
${d \epsilon_c \over dt} =  C \sigma^{m'}$.
Incorporating the constant $C$ in a redefinition of time $C t \to t$ (with
suitable redefinitions of the coefficients $\alpha/C \to \alpha$ and
$\gamma/C \to \gamma$), taking $m'=1$ and posing
$(y_1,y_2) = (\epsilon_c, \sigma)$,
we retrieve (\ref{dyn1},\ref{dyn2}).

{\bf Analysis of the dynamical system (\ref{dyn1},\ref{dyn2})
for $n>1$ and $m>1$ with $\gamma \geq 0$:}
When only the reversal term is present  (i.e., the term $y_2 |y_2|^{m-1}$ is absent),
(\ref{dyn1},\ref{dyn2})
describe a non-linear oscillator with conserved Hamiltonian
$H(\vy;n,\gamma)\equiv 
 {\gamma\over n+1}(\yo^{2})^{{n+1\over 2}}+{1\over 2}\yt^{2}$. 
Any trajectory is periodic along a constant $H$ with
period 
\be
T(H;n,\gamma) = C(n) ~ \gamma^{-1\over n+1} ~H^{{1-n\over 2(n+1)}}~,
\label{gngalla}
\ee
where $C(n)$ is a positive number that can be explicitly calculated
\cite{idesor}.
%% and bounded by
%%$|y_1| \leq [(n+1)H/\gamma]^{1\over n+1}$ and $|y_2|\leq (2H)^{1/2}$ .

When only the positive feedback is present (i.e., $\gamma=0$), 
(\ref{dyn1},\ref{dyn2}) leads to a finite-time
singularity.
The solution $\vy(t;\vyz,\tz)$ with initial condition
$\vyz=(y_{10},y_{20})$ at time $t_{0}$ can be explicitly written as:
\ba
  \yo(t;\vyz,t_{0})- y_{10} 
 &=& {\rm sign}[y_{20}] 
       \frac{(m-1)^{{m-2\over m-1}}}{2-m}~
       		  [ (t_{c}(y_{20})-t)^{{m-2\over m-1}} -
       		    (t_{c}(y_{20})-t_{0})^{{m-2\over m-1}} ] 
                ~~ \mbox{for $m\neq 2$}; \nonumber \\
 &~{\rm or}~ & {\rm sign}[y_{20}]~  
      \log \lp{t_{c}-t_{0} \over t_{c}(y_{20})-t}\rp
                ~~ \mbox{for $m=2$}, \nonumber \\
  \yt(t;\vyz,t_{0})
 &=& {\rm sign}[y_{20}]~
    (m-1)^{-{1\over m-1}}~(t_{c}(y_{20})-t)^{-{1\over m-1}} ~,
	\label{eq:sing-norm-range}
 \label{gnakal}
\end{eqnarray}
where the critical time interval
$t_{c}(y_{20})-t_{0}=\frac{1}{m-1}|y_{20}|^{1-m}$ depends only on $y_{20}$.
As $t\to t_{c}(y_{20})$, $y_2$ becomes either $+\infty$ or
$-\infty$ depending on the sign of $y_{20}$ for any $m>1$.
For $1 < m \leq 2$, it is easy to show that 
$|y_1|$ also becomes infinity as $t\to t_{c}(y_{20})$.
In contrast, for $m>2$, $y_1$ remains finite.
We thus think that $m>2$ is the relevant physical regime
for the financial, population and rupture problems discussed above.
From here on, our analysis focuses on $n>1$ and $m>2$.

\begin{figure}
\begin{center}
\includegraphics[width=18cm]{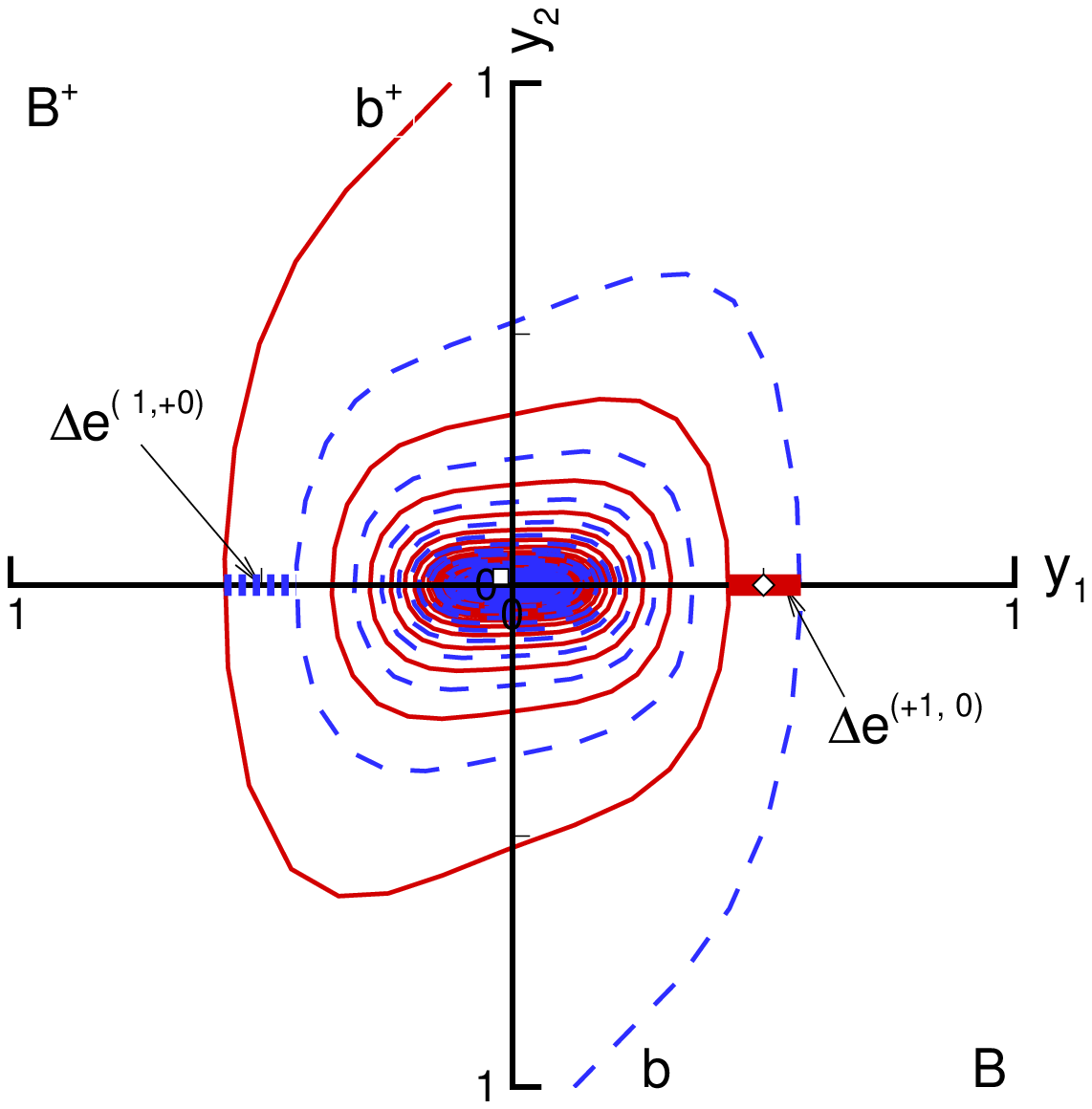}
\end{center}
\caption{Geometry of the boundaries $b^+$ and $b^-$ as
well as the basins $B^{+}$ and $B^{-}$ for
$(n,m)=(3,2.5)$ and $\gamma=10$ in phase space:
the exit segments $\triangle e^{(+1,-0)}\in B^{+}$ 
and $\triangle e^{(-1,+0)}\in B^{-}$  are thick
solid and dashed lines respectively on the $\yo$-axis; the
squares and diamonds are the initial condition and exit point of
the trajectory in Figure \ref{fg:t_n3m25g10}.}
\label{fg:basin_bBet_n3m25g10}
\end{figure}

Putting the nonlinear oscillation and positive feedback terms
together,
the dynamics of (\ref{dyn1},\ref{dyn2}) is characterized in 
figure \ref{fg:basin_bBet_n3m25g10}.
The two special intertwined trajectories along
$b^+$ (solid line) and $b^-$ (dashed line) connect
the origin $(0,0)$  to $(+\infty,+\infty)$ and $(-\infty,-\infty)$, 
respectively, and hence divide the phase space $\vy \equiv (y_1,\yt)$ 
into two distinct basins $B^{+}$ and $B^{-}$.
The basin $B^{+}$ (resp. $B^{-}$) corresponds to a finite-time
singularity $\vy_{c}(\vyz)$ with 
$y_{2c}(\vyz)=+\infty$ (resp. $y_{2c}(\vyz)=-\infty$) but 
finite $y_{1c}(\vyz)$ at the critical time $t_{c}(\vyz)$.

%%any trajectory $\vy(t;\vyz,\tz)$ with initial condition 
%%$\vyz\in B^{+}$ (resp. $\vyz\in B^{-}$)
%%will remain within it and reaches a finite-time
%%singularity, i.e., if $\vyz\in B\sppm$, then $\vy(t;\vyz,\tz)\in B\sppm$
%%for $t\in(-\infty,t_{c}(\vyz)+\tz)$ 
%%with $\lim_{t\to -\infty}\vy(t;\vyz,\tz)=(0,0)$
%%and $\lim_{t\to t_{c}(\vyz)+\tz}\vy(t;\vyz,\tz)=\vy_{c}(\vyz)$ with
%%$\vy_{c}(\vyz)=(y_{1c}(\vyz),\pm\infty)$. Here, 
%%$\vy_{c}(\vyz)$ and $t_{c}(\vyz)$ are the finite-time singularity
%%and finite-time singular interval.

Starting from $\vyz$ in $B^{+}$
close to the origin at $\tz$, a trajectory $\vy(t;\vyz,\tz)$ 
spirals out with clockwise rotation and we count a turn each time it
crosses the $\yo$-axis,  
i.e., $\yo$ changes its direction of motion ($d\yo/dt=0$).
Deep in the spiral structure, $\vy(t;\vyz,\tz)$
follows approximately the orbit of constant Hamiltonian 
$H$ defined by the nonlinear oscillator
but fails to close on itself  because
$H$ is no longer conserved due to the positive feedback:
\be
{d\over dt}H (\vy;n,m,\gamma) = |\yt|^{m+1}\geq 0~.
\label{eq:all-phase-dHdt}
\ee
This slowly growing nonlinear oscillator defines the first dynamical
regime.

\begin{figure}
\begin{center}
\includegraphics[height=16cm]{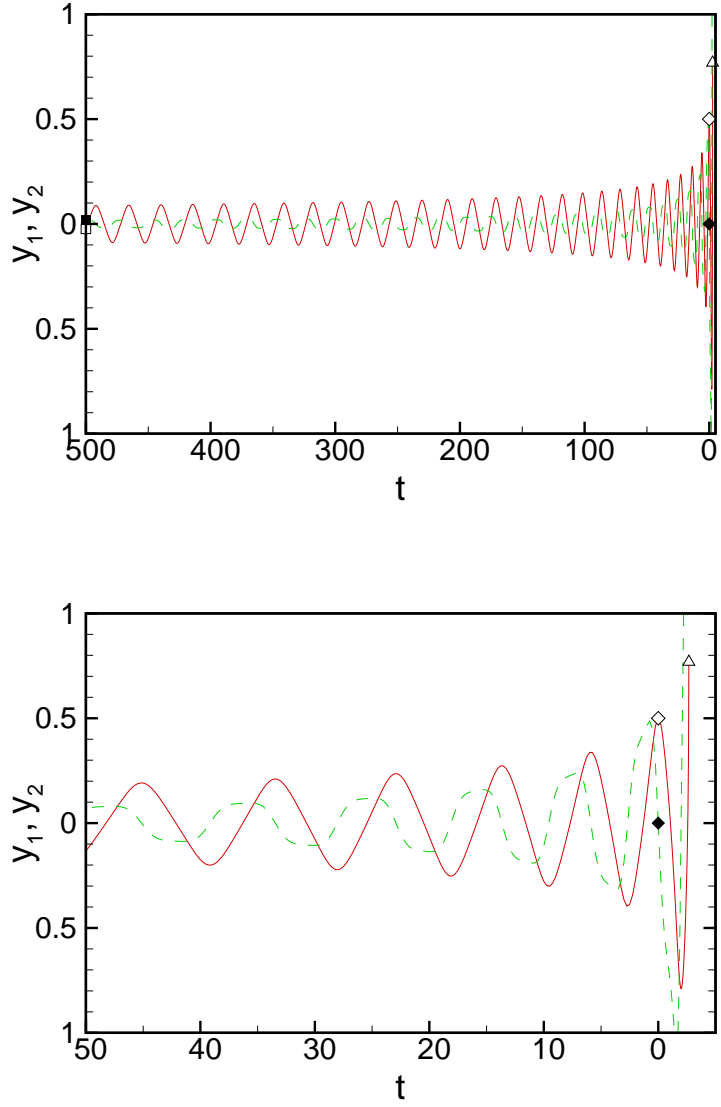}
\end{center}
\caption{A trajectory in $B^{+}$ starting
 from initial condition
$(-2.32878\times 10^{-2},1.71083\times 10^{-2})$ at $t=-500$,
going through the exit point
$(0.5,0)$ at $t=0$, to the critical point
$(0.768899,+\infty)$ at $t_{c}=2.67284$:
Solid and dashed lines correspond to $y_{1}$ and $y_{2}$, respectively;
square, diamond and triangle are
the initial condition, the point in the exit segment
$\triangle e^{(+1,-0)}$, and the critical point along the trajectory with 
open and filled symbols corresponding to $y_{1}$ and $y_{2}$.}
\label{fg:t_n3m25g10}
\end{figure}

Any trajectory starting in the first dynamical regime must eventually 
cross-over to the second one associated with a route
to the singularity without any further oscillation.
The second dynamical regime occurs $y_2$ diverges (and therefore $dy_2/dt$ also diverges)
while $y_1$ remains finite on the approach to $t_c(\vyz)$. As a consequence,
the reversal term $\gamma y_1|y_1|^{n-1}$ in $d\yt/dt$ (\ref{dyn2}) 
becomes negligible close to $t_c(\vyz)$ and 
(\ref{gnakal}) is the asymptotic solution of
(\ref{dyn1},\ref{dyn2}). 
Figure \ref{fg:t_n3m25g10} shows a typical time series of a 
$\vy(t;\vyz,\tz)$ starting from $\vyz$ near the origin.

In the basin $B^{+}$, the transition from the 
first to second dynamical regime occurs at the
exit segment $\triangle e^{(+1,-0)}$ 
on the positive $\yo$-axis (figure \ref{fg:basin_bBet_n3m25g10})
whose right and left end-points 
according to the forward direction of the flow
are defined by the boundaries $b^{+}$ and $b^{-}$, respectively.
In forward time, $\triangle e^{(+1,-0)}$ fans out rapidly over 
$\yo\in(-\infty,\infty)$ as it reaches a singularity with 
$\yt\to +\infty$ in the second dynamical regime.
Note that $|\yo|$ can reach $\infty$ if and only if 
$\vyz$ is at an end point  of $\triangle e^{(+1,-0)}$, 
i.e., on  either $b^{+}$ or $b^{-}$.
Similar results hold for the exit segment $\triangle e^{(-1,+0)}$ in $B^{-}$
but with $\yt\to -\infty$.
In backward time, $\triangle e^{(+1,-0)}$ 
and $\triangle e^{(-1,+0)}$ swirl into the origin
while making countable infinite many turns.
These backward swirls of the exit segments completely define
the first dynamical regime.

\begin{figure}
\begin{center}
\includegraphics[width=18cm]{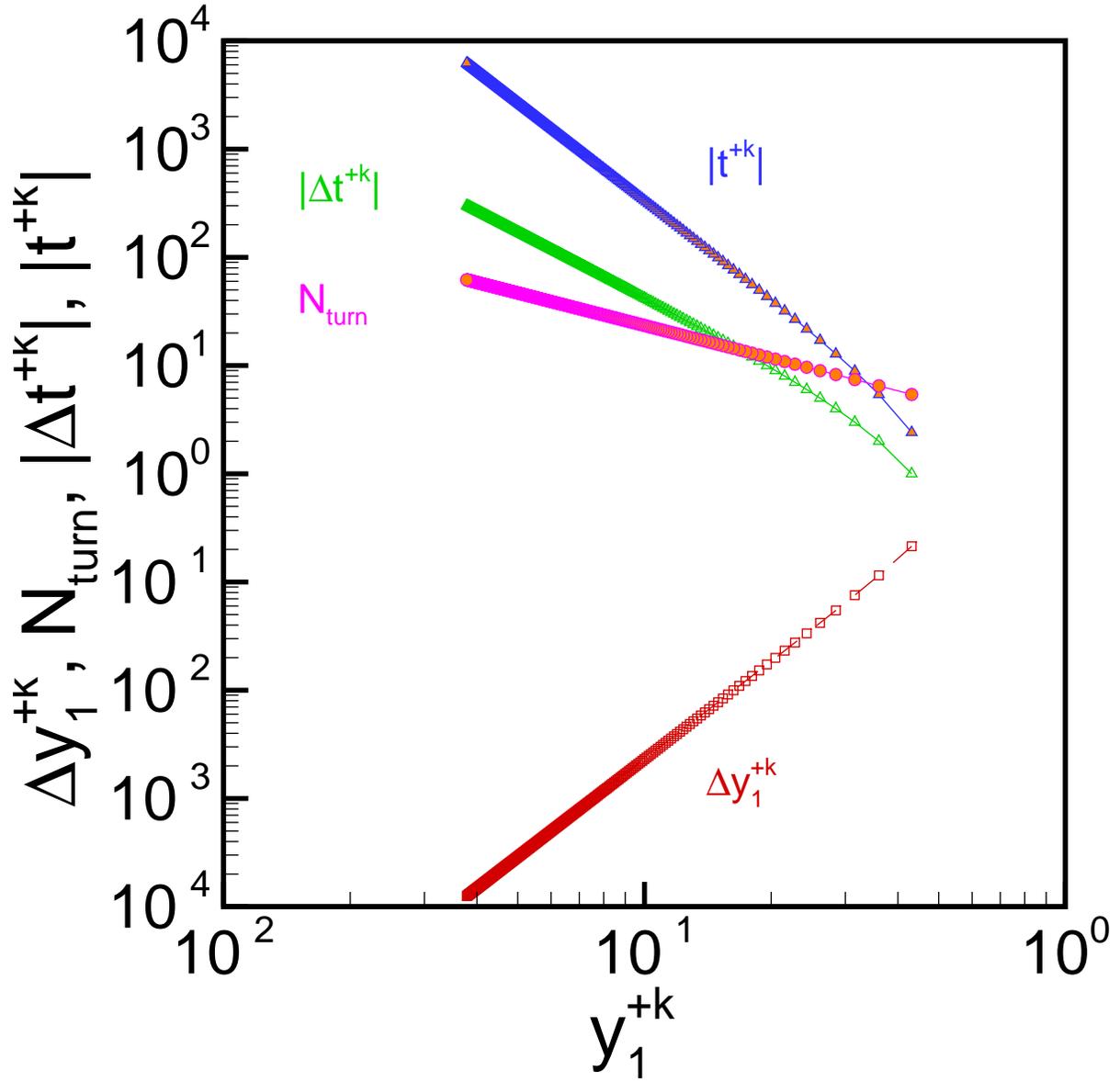}
\end{center}
\caption{Scaling laws associated with the self-similar properties
of the nonlinear oscillatory regime
as a function of initial conditions at
turn points for $(n,m)=(3,2.5)$ and  $\gamma=10$.}
\label{fg:s_n3m25g10}
\end{figure}

The first dynamical regime exhibits
remarkable scaling properties that 
we quantify by the dependence on 
the initial condition $\vyz=(y_{10},0)$ on the $\yo$-axis of the following
quantities: the
number ($N_{turn})$ of turns before reaching the singularity,
the exit time ($t_{e}$) to reach the exit segment into
the second dynamical regime, the
time interval ($\Delta t_{e}$) and the
increment ($\Delta \yo$) in the amplitude of $\yo$ 
over one turn.
%%$N_{turn}$ is directly associated with the oscillatory behavior of the
%%dynamics. $y_{1c}$ is a property more directly associated with the singular
%%behavior at the end of the dynamics
%%The critical time $t_{c}$ can be decomposed as the sum of the
%%duration of the  
%%first oscillatory regime to reach the exit segment
%%and of the second singular regime, but is mostly dominated by the former. 
Figure \ref{fg:s_n3m25g10} shows log-log plots of the scaling
properties measured at the $k$-th backward intersection of
$b^{+}$ with the $\yo$-axis starting from $k=0$ at the out-most
intersection (the left end-point of $\triangle e^{(-1,+0)}$
in figure \ref{fg:basin_bBet_n3m25g10}).
In order to get accurate and reliable
estimations of these dependences and of the exponents
defined below, we have integrated the dynamical
equations using a fifth-order Runge-Kutta integration 
scheme with adjustable time step.

The log-log dependences shown in figure \ref{fg:s_n3m25g10}
qualify power laws defined by
\ba
N_{\rm turn} &\sim& y_{10}^{-a}~,~~~~~~~~{\rm where}~ a>0~,  \label{rel1}\\
t_{e} &\sim& y_{10}^{-b}~,~~~~~~~~{\rm where}~ b>0~, \label{rel2}\\
\Delta y_{10} &\sim& y_{10}^{c}~,~~~~~~~~{\rm where}~ c>0~, \label{rel3}\\
\Delta t_{e} &\sim& y_{10}^{-d}~,~~~~~~~~{\rm where}~ d>0~. \label{rel4}
\ea
Eliminating $y_{10}$ between (\ref{rel2}) and (\ref{rel4}) gives
$\Delta t_{e} \sim t_{e}^{d \over b}$.
Since $\Delta t_{e}$ is nothing but the difference
$\Delta t_{e}= t_{e}(N_{\rm turn}+1) - t_{e}(N_{\rm turn})$, this
gives the discrete difference equation
${t_{e}(N_{\rm turn}+\Delta N_{\rm turn})
  - t_{e}(N_{\rm turn}) \over \Delta N_{\rm turn}}
 \propto t_{e}^{d \over b}$, where $\Delta N_{\rm turn}
 = N_{\rm turn}+1 - N_{\rm turn}=1$. 
This provides a discrete difference representation of the 
derivative $dt_{e}/dN_{\rm turn}$ which can be integrated
formally to give $N_{\rm turn} \sim t_{e}^{1-{d \over b}}$, which is
valid for $d<b$. Comparing with the
relation between $N_{\rm turn}$ and 
$t_{e}$ obtained by eliminating $y_{10}$ between (\ref{rel1}) and  (\ref{rel2}),
i.e., $N_{\rm turn} \sim t_{e}^{a \over b}$, we get the scaling relation
\be
a = b-d~.  \label{scal1}
\ee
Since $a>0$, the condition $d<b$ is automatically verified.

Similarly, 
$\Delta y_{10} = y_{10}(N_{\rm turn}+1) - y_{10}(N_{\rm turn}) 
 \propto y_{10}^{c}$ according to definition (\ref{rel3}). 
This gives the differential equation 
$dy_{10}/dN_{\rm turn} \sim y_{10}^c$, whose solution is $N_{\rm turn} 
\sim 1/y_{10}^{c-1}$, valid for $c>1$. 
Comparing with the definition (\ref{rel1}), we get the second scaling relation
\be
a = c-1~.  \label{scal2}
\ee
Since $a>0$, the condition $c>1$ is automatically satisfied.

Deep in the spiral structure depicted in 
figure \ref{fg:basin_bBet_n3m25g10}, in the presence
of the positive feedback term, one rotation around the origin is not exactly
closed but the failure 
to close, which is very small especially near the origin, is quantified by 
$\Delta y_{10}$ shown in figure 2 which follows (\ref{rel3}).
We approximate the time
$\Delta t_{e}$ needed to make one full (almost closed) rotation by
the period $T(H)$ without the positive feedback term. 
This is essentially an adiabatic approximation in which the Hamiltonian
$H$ and the period $T(H)$ are assumed to vary sufficiently slowly so that
the local period of rotation follows adiabatically the variation of the
Hamiltonian $H$. Putting together (\ref{gngalla}) and the 
fact that the amplitude of
$y_{10}$ is proportional to $H^{1 \over n+1}$, we get
$\Delta t_{e} \sim |y_{10}|^{1-n \over 2}$, 
which, by comparison with (\ref{rel4}), gives
\be
d = {n-1 \over 2}~.  \label{eqexp1}
\ee

The last equation is provided by $dT/dt$, expressed as 
$dT/dt=(dT/dH) \times (dH/dt)$, where 
$dT/dH$ is obtained by differentiating (\ref{gngalla}) and $dH/dt$ is given by
(\ref{eq:all-phase-dHdt}). Estimating $dT/dH$ from (\ref{gngalla})
is consistent with the above approximation in which a full rotation
along the spiral takes the same time as the corresponding closed orbit
in absence of the positive feedback term. Expressing $dH/dt$ using
(\ref{eq:all-phase-dHdt}) involves another approximation, which is
similar in spirit to a mean-field approximation corresponding to average
out the effect of the positive feedback term over one full rotation. In so doing, we
average out the subtle positive and negative interferences
between the reversion and positive feedback terms.
We replace $dT/dt$ by $d \Delta t_{e} / d t_{e}$ and obtain
${d \Delta t_{e} \over d t_{e}} \sim H^{-{3n+1 \over 2(n+1)}}~ |\yt|^{m+1}
\sim \Delta t_{e}^{3n+1 \over n-1}~ |y_{10}|^{(n+1)(m+1) \over 2}$,
where the dependence in $\Delta t_{e}$ is derived by replacing 
$H$ by its dependence  as a function of $T$ (by inverting $T(H)$) 
and by identifying $T$ and $\Delta t_{c}$. 
We have also used $|\yt| \sim |y_{10}|^{n+1 \over 2}$.
Taking the derivative of $\Delta t_{e} \sim t_{e}^{d \over b}$ 
with respect to $t_{e}$ provides another estimation of ${d \Delta t_{e} \over d t_{e}}$,
and replacing
$\Delta t_{e}$ in the above equation by its dependence as a function of $y_{10}$
as defined by (\ref{rel4}) gives finally:
\be
a = {1 \over 2} (n+1)(m+1) - {1 \over 2} (3 n +1)~.  \label{eqexp2}
\ee
We find a perfect agreement between the 
theoretical predictions 
(\ref{scal1}), (\ref{scal2}), (\ref{eqexp1}), (\ref{eqexp2})
for the exponents $a, b, c, d$ defined by (\ref{rel1})-(\ref{rel4})
with an estimation obtained from the direct numerical integration of
the dynamical equations \cite{idesor}. We have
also verified the independence of the exponents 
$a, b, c, d$ with respect to the amplitude $\gamma$ of the reversal term.

In the oscillatory regime, the growth of the amplitude
of $y_1(t)$ follows a power law similar to (\ref{gnakal}). We obtain
it by combining some of the previous scaling laws (\ref{rel1}-\ref{rel4}). 
Indeed, taking the ratio of (\ref{rel3}) and (\ref{rel4}) yields
$\Delta y_1/\Delta t_{e} \sim y_1^{c+d}$. Since $\Delta y_1$ corresponds to
the growth of the local amplitude $A_{y_1}$ of the oscillations of
$y_1(t) $ due to the positive feedback term over one turn of the spiral
in phase space, this turn lasting $\Delta t_{e}$,
we identify this scaling law with the equation for the growth rate of 
the amplitude $A_{y_1}$ of $y_1$ in this oscillatory regime:
${d A_{y_1} \over dt} \sim A_{y_1}^{c+d}$ whose solution is
$A_{y_1}(t) = {B \over (t^*-t)^{1/b}}$,
where $B$ is another amplitude. This prediction is verified accurately from our direct
numerical integration of the equations of motion \cite{idesor}.
We have used the scaling relations (\ref{scal1}) and (\ref{scal2}) leading 
to $c+d-1=b$. The time $t^*$ is a constant of integration such that 
$B/(t^*)^{1 \over b}=A_{y_1}(t_0)$, which can be interpreted as an
{\it apparent} or ``ghost'' critical time.
$t^*$ has no reason to be equal to $t_{e}$, in particular since the extrapolation
of $A_{y_1}(t) = {B \over (t^*-t)^{1/b}}$
too close to $t^*$ would predict a divergence of $y_1(t)$.
The dynamical origin of the difference between $t^*$ and $t_c$ comes from the fact 
that $t^*$ is determined by the oscillatory regime while $t_c$
is the sum of two contributions, one from
the oscillatory regime and the other from the singular regime.

Combining (\ref{rel4}) with this solution for $A_{y_1}(t)$
gives the time dependence of the
local period $\Delta t_{e}$ of the oscillation in the oscillatory regime as
$\Delta t_{e} = t_{k+1} - t_k \sim (t^*-t_k)^{d \over b}$, where $k$ is the turn index 
previously used. This result generalizes 
the log-periodic oscillation associated with discrete scale invariance (DSI)
characterized by $t_{k+1} - t_k \sim 1/\lambda^k$ 
(where $\lambda>1$ is a preferred scaling ratio of DSI), which is
recovered in the limit $d/b \to 1^-$ (corresponding to 
($n \to \infty, m \to 2$). The dynamical system (\ref{dyn1},\ref{dyn2}) provides
a mechanism for generalized
log-periodic oscillations with, in addition, a finite number of them 
due to the cross-over to the non-oscillatory regime. We shall report elsewhere
on tests of this theory on financial and rupture data.

{\bf Acknowledgments:} This work was partially supported by 
ONR N00014-99-1-0020 (KI) and by
NSF-DMR99-71475 and the James S. Mc Donnell Foundation 21st century 
scientist award/studying complex system (DS).

\pagebreak

\pagebreak

\pagebreak
\end{document}